\documentclass[aps,prl,preprintnumbers,twocolumn,superscriptaddress,nofootinbib,floatfix,10pt]{revtex4}
\usepackage{amsfonts,amssymb,stmaryrd,latexsym,amsmath,braket}
\usepackage{graphicx,mathtools}
\usepackage[caption=false]{subfig}
\usepackage{times}
\usepackage{slashed}
\usepackage{braket}
\usepackage{comment}
\usepackage[utf8]{inputenc}
\newcommand{\beginsupplement}{%
        \setcounter{table}{0}
        \renewcommand{\thetable}{S\arabic{table}}%
        \setcounter{figure}{0}
        \renewcommand{\thefigure}{S\arabic{figure}}%
        \setcounter{equation}{0}
        \renewcommand{\theequation}{S\arabic{equation}}        
     }
\begin{document}
        
\title{Self-Learning Emulators and Eigenvector Continuation}

\author{Avik Sarkar}
\email{sarkarav@msu.edu}
\affiliation{Facility for Rare Isotope Beams and Department of Physics and Astronomy,
Michigan State University, East Lansing, MI 48824, USA}

\author{Dean Lee}
\email{leed@frib.msu.edu}
\affiliation{Facility for Rare Isotope Beams and Department of Physics and Astronomy,
Michigan State University, East Lansing, MI 48824, USA}
        
        \begin{abstract}
        Emulators that can bypass computationally expensive scientific calculations with high accuracy and speed can enable new studies of fundamental science as well as more potential applications.  In this work we discuss solving a system of constraint equations efficiently using a self-learning emulator.  A self-learning emulator is an active learning protocol that can be used with any emulator that faithfully reproduces the exact solution at selected training points.  The key ingredient is a fast estimate of the emulator error that becomes progressively more accurate as the emulator is improved, and the accuracy of the error estimate can be corrected using machine learning.  We illustrate with three examples.  The first uses cubic spline interpolation to find the solution of a transcendental equation with variable coefficients. The second example compares a spline emulator and a reduced basis method emulator to find solutions of a parameterized differential equation. The third example uses eigenvector continuation to find the eigenvectors and eigenvalues of a large Hamiltonian matrix that depends on several control parameters.
        \end{abstract}
\maketitle

\paragraph*{Introduction} The frontiers of scientific discovery often reside just beyond the limits of computability.  This explains the great interest across many scientific disciplines in using machine learning tools to build efficient emulators that predict scientific processes beyond what is possible with direct calculations \cite{Carleo:2019,Thiagarajan:2020,Kasim:2020,Bedaque:2020pct}.  However, a problem arises in that large amounts of training data for such an emulator are not possible since the required computations are difficult and expensive.  In this work, we provide a potential solution to this problem when the objective is to solve a system of constraint equations over some domain of control parameters. We introduce a method called self-learning emulation, an active learning protocol \cite{settles.tr09,cohn1996active,cohn1994improving} that relies on a fast estimate of the emulator error and a greedy local optimization algorithm that becomes progressively more accurate as the emulator improves.  Provided that the emulator faithfully reproduces the exact solution at the training points, the error will decrease with the number of training points as either a power law for piecewise continuous emulators or exponentially fast for smooth function emulators.  The resulting acceleration is typically several orders of magnitude or more, and the gain in computational speed is achieved by using the emulator itself to estimate the error.  As we will show, self-learning emulators are highly efficient algorithms that offer both high speed and accuracy as well as a reliable estimate of the error.  We note that the self-learning emulators we discuss here are qualitatively different from other machine learning algorithms that model the solutions using gradient descent optimization of some chosen loss function.  While these gradient descent optimization methods are highly parallelizable and very fast, they usually suffer from critical slowing down with respect to error and cannot achieve arbitrarily high accuracy in polynomial computing time.  Sometimes scientific discovery requires seeing very small but important new phenomena that might otherwise be absent in approximate machine learning models.

We will demonstrate several contrasting examples of self-learning emulators.  The first uses a cubic spline emulator to find the solution of a transcendental equation with variable coefficients. The second example uses the spline emulator and a reduced basis method emulator to find solutions of a parameterized differential equation. The third example is our primary example for quantum many body calculations.  It uses eigenvector continuation to find the eigenvectors and eigenvalues of a large Hamiltonian matrix that depends on several control parameters.  See Ref.~\cite{Frame:2017fah,Frame:2019jsw} for an introduction to eigenvector continuation and Ref.~\cite{Konig:2019adq,Ekstrom:2019lss} for applications to the quantum many body problem.

\paragraph*{Constraint equations and error estimates} We consider a general set of simultaneous constraint equations $G_i({\bf x},{\bf c})=0$ that we solve for variables ${\bf x}=\{x_j\}$ as a function of control parameters ${\bf c}=\{c_k\}$ over some domain ${\bf D}$.  Let us denote the exact solutions as ${\bf x}({\bf c})$.  We assume that we have an emulator which can take the exact solutions for some set of training points $\{{\bf c}^{(i)}\}$ and construct an approximate solution ${\bf \tilde x}({\bf c})$ for all ${\bf c}\in{\bf D}$.  Let us define the error or loss function as the norm $\lVert \Delta{\bf x}({\bf c}) \rVert $ of the residual $\Delta{\bf x}({\bf c}) = {\bf x}({\bf c})-{\bf \tilde x }({\bf c})$.  The objective is to train the emulator to minimize the peak value of the error function over the domain ${\bf D}$ using as few additional training points as possible.

Since the error function will vary over many orders of magnitude, it is more convenient to work with the natural logarithm of the error function, $\log \lVert \Delta{\bf x}({\bf c}) \rVert $. The emulator will reproduce the exact solution at the training points $\{{\bf c}^{(i)}\}$.  Therefore, the logarithm of the error function will become a rapidly varying function of ${\bf c}$ as we include more training points.  

Let us consider the case where $\Delta{\bf x}({\bf c})$ is small enough that we can accurately expand the constraint equations as 
\begin{equation}
 G_i({\bf \tilde x}({\bf c}),{\bf c}) + \Delta {\bf x}({\bf c}) {\bf \cdot} {\bf \nabla_{\bf x}}G_i({\bf \tilde x}({\bf c}),{\bf c}) \approx 0. \label{Eq:Linear_error_constraint}
\end{equation}
If the number of degrees of freedom is small, we can solve the linear inversion problem for $\Delta {\bf x}({\bf c})$ and provide a fast estimate for the logarithm of the error. This estimate is nothing more than the multivariate form of the Newton-Raphson method.  

\paragraph*{Fast error estimates} For most cases of interest, however, there will be many degrees of freedom and the matrix inversion required to solve for $\Delta {\bf x}({\bf c})$ will be too slow for our self-learning emulator training process.  We therefore choose another non-negative functional $F[\{G_i({\bf \tilde x}({\bf c}),{\bf c})\}]$ as a surrogate for $\lVert \Delta{\bf x}({\bf c}) \rVert $.  The only essential requirement we impose on $F[\{G_i({\bf \tilde x}({\bf c}),{\bf c})\}]$ is that it is linearly proportional to $\lVert \Delta{\bf x}({\bf c}) \rVert $ in the limit $\lVert \Delta{\bf x}({\bf c}) \rVert \rightarrow 0$.  This allows us to write the logarithm of the error as
\begin{align}
    \log \lVert \Delta{\bf x}({\bf c}) \rVert = \log F[\{G_i({\bf \tilde x}({\bf c}),{\bf c})\}] + A + B({\bf c}), \label{quick_error}
\end{align}
where $A$ is a constant and the average of $B({\bf c})$ over the domain ${\bf D}$ is zero.  Since $F[\{G_i({\bf \tilde x}({\bf c}),{\bf c})\}]$ is linearly proportional to $\lVert \Delta{\bf x}({\bf c}) \rVert $ in the limit $\lVert \Delta{\bf x}({\bf c}) \rVert \rightarrow 0$, the function $\log F[\{G_i({\bf \tilde x}({\bf c}),{\bf c})\}]$ will have the same steep hills and valleys as the function $\log \lVert \Delta{\bf x}({\bf c}) \rVert$ as we include more training points.  In the limit of large number of training points, we can neglect the much smaller variation of $B({\bf c})$ over the domain ${\bf D}$.  We can therefore approximate the logarithm of the error as $\log F[\{G_i({\bf \tilde x}({\bf c}),{\bf c})\}] + A$.  We note that the unknown constant $A$ is irrelevant for comparing the logarithm of the error for different points ${\bf c}$.  Nevertheless, we can also quickly estimate $A$ simply by taking several random samples of ${\bf c}$ and computing the average value of the difference between $\log \lVert \Delta{\bf x}({\bf c}) \rVert $ and $\log F[\{G_i({\bf \tilde x}({\bf c}),{\bf c})\}]$. We can refine this estimate further using machine learning to approximate the function $B({\bf c})$. In several of our examples we show the improvement resulting from these additional steps. 

The self-learning emulator training program is a greedy algorithm where we search to find the point ${\bf c}$ where the logarithm of the error is greatest.  We then add this point to the training set and repeat the whole process.   In this manner we have constructed a fast emulator that becomes more and more accurate as more training points are added and provides a reliable estimate of the emulator error.  It should be emphasized that the self-learning emulation is just an algorithm to learn the best training points for the emulator, and it does not change the process of emulation itself. Thus it can be used with any emulator that faithfully reproduces the exact solution at the training points. This could be a simple method such as polynomial interpolation or a Gaussian process, or a more involved method such as neural networks or eigenvector continuation. We retain all the beneficial properties of the emulator such as its computational speed advantage, parallelizablilty, ease of application, etc. It can be applied to any system of constraints such as solutions of algebraic or transcendental equations, linear and nonlinear differential equations, and linear and nonlinear eigenvalue problems.  

\paragraph*{Model 1} For the first example, Model 1, we use a natural cubic spline emulator to find the lowest real solution of a transcendental equation.  We consider the solution to the equation 
\begin{align}
    c_5 x^5 + c_4 x^4\sin (10x) + c_3 x^3 + c_2 x^2 + c_1 x + c_0 = 0, \label{Eq:Sin_polynomial}
\end{align}
where all the coefficients $c_i$ are real. We fix coefficients $c_5=c_3=c_2=c_1=c_0=1$, and we vary the coefficient $c_4$. We are interested in the lowest real $x$ that satisfies Eq.~(\ref{Eq:Sin_polynomial}). We know that a real solution for $x$ always exist for real $c_4$, however the dependence of the solution on $c_4$ is not trivial and has discontinuities with respect to parameter $c_4$.

We start with three training points for $c_{4}$, two on the boundary and one in the interior, and use natural cubic splines to define the cubic spline approximation ${\tilde x}(c_{4})$ for all values of $c_{4}$.  The logarithm of the error function is then $\log \vert \Delta x(c_{4}) \vert $ where $\Delta x(c_{4})= x(c_{4})-{\tilde x}(c_{4})$. 
We can estimate $\vert \Delta x(c_{4}) \vert$ using the Newton-Raphson method, 
\begin{align}
    \vert \Delta x(c_{4}) \vert \approx \frac{\lvert p({\tilde x}(c_{4})) \rvert }{\sqrt{\lvert p'({\tilde x}(c_{4}))\rvert ^2+\epsilon^2}}, \label{regulated}
\end{align}
where we have included a small regulator $\epsilon$ to avoid divergences when the derivative $p'$ vanishes. We use the right-hand side of Eq.~(\ref{regulated}) for our error estimate with $\epsilon = 1$.

\begin{figure}[hbt]
     % \centering
    \includegraphics[width=8.4cm]{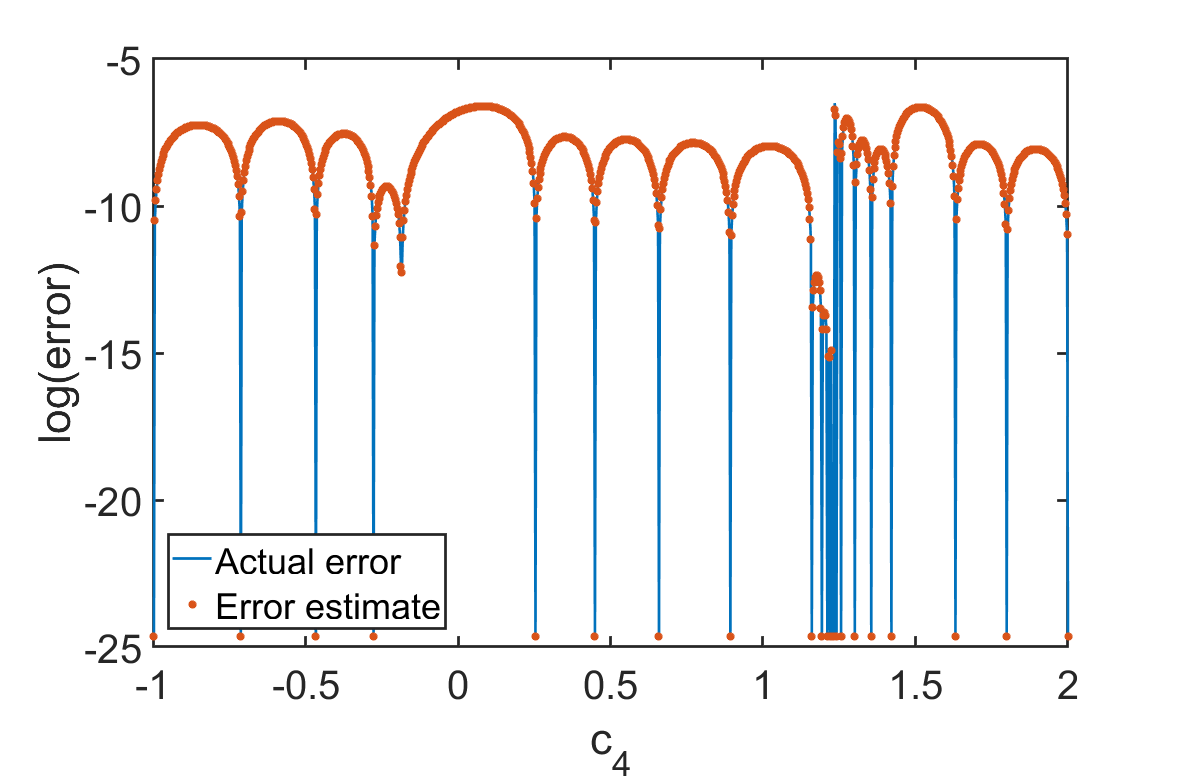}\\
    \caption{Logarithm of the actual error and error estimate for the cubic spline self-learning emulator in Model 1 after $20$ iterations.}
    \label{Fig:spline_2}
\end{figure}

In Fig.~\ref{Fig:spline_2} we show results for the logarithm of the error estimate and actual error, spanning the interval from $c_4 = -1$ to $c_4=2$ with $23$ training points.  The fact that more training points are needed near $c_4\approx1.2$ shows that the training process is not simply adding more training points at random, but is instead uniformly improving the emulator performance across the entire domain.   As shown in the Supplemental Material, there is a discontinuity at $c_4\approx1.232$, and we need a higher density of training points near the discontinuity.   Fig.~\ref{Fig:spline_2} shows that our error estimates are matching well with the actual error. Therefore both $A$ and $B({\bf c})$ as defined in Eq.~(\ref{quick_error}) are negligible for Model 1.

In the limit of large number of training points, $N$, the error for the spline interpolation for a smooth function scales as $O(N^{-4})$ \cite{Ahlberg_1967}.  For the case of Model 1, however, the exact solution has a jump discontinuity, and so the power law scaling is slower.  Numerically, we find that the error is approximately $O(N^{-2.2})$.  See the Supplemental Material for details on the error scaling versus number of training points as well as the dependence on the choice of initial training points. On a single Intel i7-9750H processor, evaluating the exact solution using standard root finding methods for one value of $c_4$ requires about $10^{-1}$~s of computational time.  In contrast, it takes about $ 10^{-6}$~s for spline interpolation for $23$ training points.  The raw emulator speedup factor is therefore $s_{\rm raw} \sim 10^{5}$.  Let $M$ be the number of evaluations of needed and $N_{\epsilon}$ be the number of emulator training points needed to achieve error tolerance $\epsilon$. The overall computational speedup factor for the self-learning emulator can then be estimated by the minimum of $M/N_{\epsilon}$ and $s_{\rm raw}$. If the fast error estimate were not used, then $N_{\epsilon}$ would be replaced by the number of evaluations needed to train the emulator to the desired error tolerance $\epsilon$, which is generally much larger than $N_{\epsilon}$. 

%--------------------Model 2: Differential Equation--------------------

\paragraph*{Model 2} In our next example, Model 2, we will emulate the solution of an ordinary differential equation with one variable $z$ and one control parameter $c$.  We consider a family of differential equations $L x(z) = 0$, where
\begin{align}
    L = \frac{1}{(1+2z)^2} \frac{d^2}{dz^2}-\frac{2}{(1+2z)^3}\frac{d}{dz}  + c^2e^{2c}, 
\end{align}
and $c$ is a real parameter. Our boundary conditions are $x(z=0,c) = 0$ and $\partial_z x(z=0,c) = 1$ for all $c$. We consider the region $0\leq z\leq 1$, and $0\leq c\leq 1$. The exact solution is $x(z,c) = \frac{1}{ce^c}\sin[ce^c(z+z^2)]$.
We consider two different emulators.  The first is the natural spline emulator, which we use to perform interpolations and extrapolations in $c$ for each value of $z$.  The second emulator is a reduced basis emulator, which uses high-fidelity solutions of the differential equation for several training values of $c$ and solves the constraint equations approximately using subspace projection.  Reduced basis (RB) emulators have proven useful for solving computationally-intensive parameterized partial differential equations \cite{Bonilla:2022rph,Melendez:2022kid,Reduced_basis_book, Quarteroni_2011, Field_2011}. 

For our fast error estimate $F[\tilde{x}(z,c),c]$, we need some function that is linearly proportional to the actual error $\lVert \Delta{x}({z,c}) \rVert$ in the limit $\lVert \Delta{x}({z,c}) \rVert \rightarrow 0$. There are many good choices one can make, and here we choose
\begin{align}
    F[\tilde{x}(z,c),c] = \left\Vert \frac{L \Tilde{x}(z,c)}{\sqrt{ \big(\frac{d}{dz}L\tilde{x}(z,c)\big)^2 +\epsilon^2}} \right\Vert_1,
\end{align}
where we have again included a small regulator $\epsilon$ to avoid divergences.  Here we are using the $L_1$ norm, which is the integral over $z$ of the absolute value.

\begin{figure}
     % \centering
    \includegraphics[width=8.4cm]{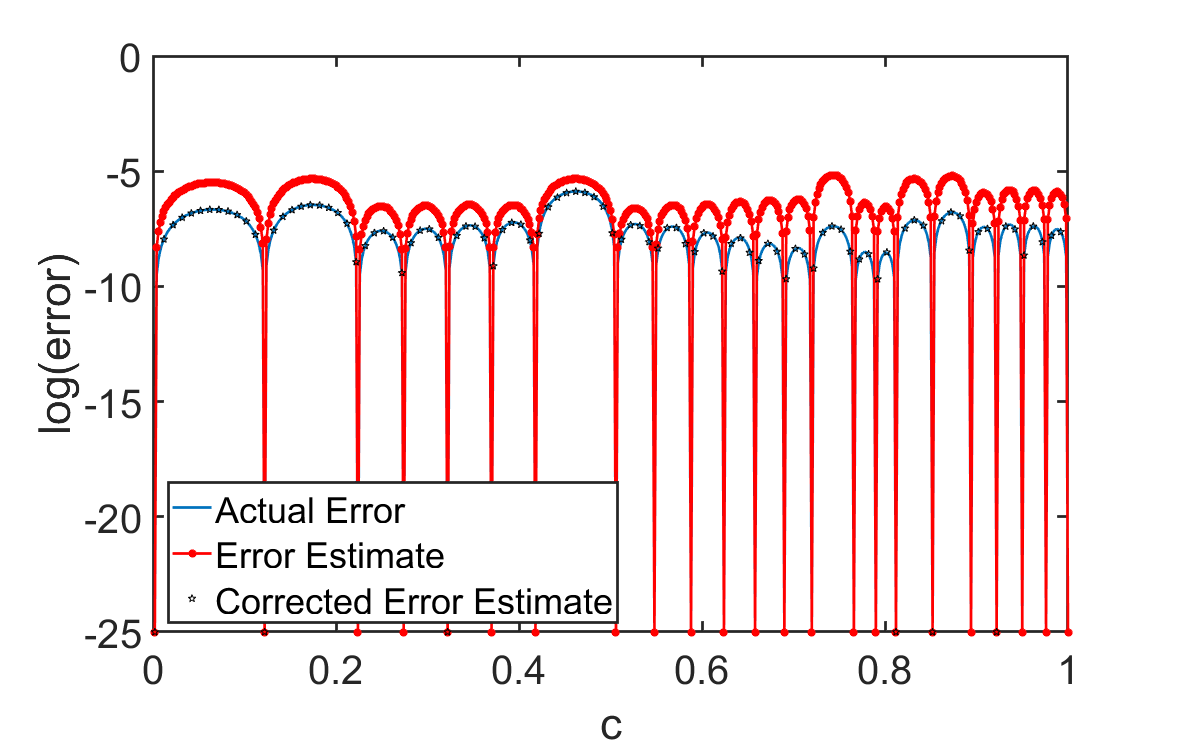}\\
    \caption{Logarithm of the actual error, error estimate, and corrected error estimate for the natural spline emulator with self-learning in Model 2 after $20$ iterations.}
    \label{Fig:Spline_ode_result}
\end{figure}

\begin{figure}
     % \centering
    \includegraphics[width=8.4cm]{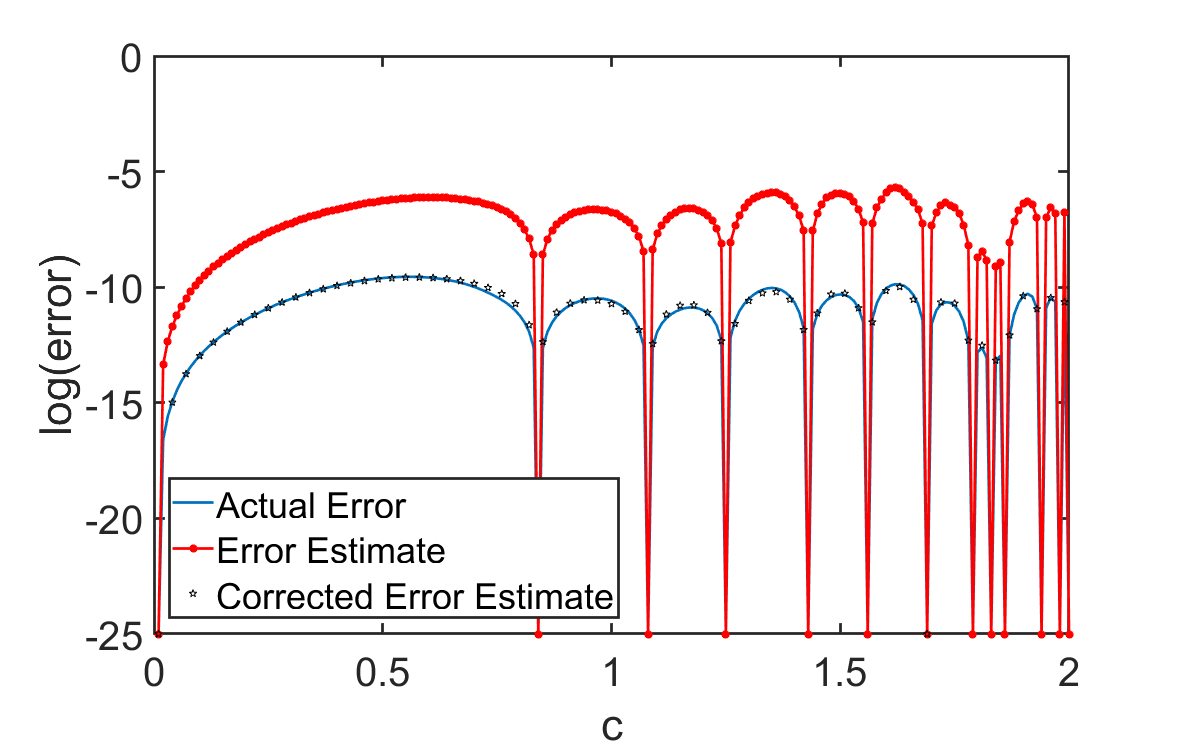}\\
    \caption{Logarithm of the actual error, error estimate, and corrected error estimate for the reduced basis emulator with self-learning in Model 2 after $10$ iterations.}
    \label{Fig:Basis_method_ode_result}
\end{figure}

We initialize with two training points at the boundaries and one in the interior.  For the spline emulator, the error scales approximately as $O(N^{-1.88})$ for $N$ up to $300$ training points.  Meanwhile, the error for the RB emulators scales exponentially fast in $N$.  We have therefore extended the domain to the wider interval $0\leq c\leq 2$ in order to show more details of the performance before reaching the limits of machine precision.  Over this interval, the RB emulator error scales approximately as $O(e^{-2.66N})$, for $N$ above $10$ training points.   Fig.~\ref{Fig:Spline_ode_result} shows the actual error and estimated error after 20 iterations of the self-learning spline emulator. Fig.~\ref{Fig:Basis_method_ode_result} shows the actual error and estimated error after 10 iterations of the self-learning RB algorithm.  In both cases the difference between the actual error and estimate error is a slowly-varying function of $c$ as predicted.  We note that the exact solution $x(z,c) = \frac{1}{ce^c}\sin[ce^c(z+z^2)]$ oscillates more rapidly with increasing $c$, and the emulators therefore need more training points for larger $c$.  

We can estimate the difference between the error estimate and the actual error by constructing a Gaussian Process (GP) emulator for the difference function $A + B(c)$.  We train the GP by computing $A + B(c)$ at the midpoints in between the emulator training points.  We have performed this correction for both the spline and RB emulators, and the results are shown in Figs.~\ref{Fig:Spline_ode_result} and \ref{Fig:Basis_method_ode_result}.  We see that the corrected error estimate is in excellent agreement with the actual error.

On a single Intel i7-9750H processor, numerically solving the differential equation for one value of $c$ takes about $7\times 10^{-2}$~s.  In contrast the spline emulator requires about $1.7\times 10^{-3}$~s for $23$ training points, and the RB emulator takes about $5.5\times 10^{-4}$~s for $13$ training points. Therefore the spline emulator has a raw speedup factor of $s_{\rm raw} \sim 40$, while the RB emulator has a raw speedup factor of $s_{\rm raw} \sim 130$.  Given the somewhat comparable values for $s_{\rm raw}$ and the exponential scaling of the error for the RB emulator, we conclude that the RB emulator significantly outperforms the spline emulator for this example.

%--------------------Model 3: EC--------------------

\paragraph*{Model 3} For our final example, Model 3, we will use eigenvector continuation as the emulator.  Eigenvector continuation (EC) belongs to the family of RB methods \cite{Bonilla:2022rph,Melendez:2022kid}, however the applications may involve extremely large vector spaces where general vector operators may not be possible \cite{Frame:2017fah}.  EC consists of projecting the Hamiltonian onto a subspace spanned by a set of exact eigenvectors of the Hamiltonian for selected training points and then solving the resulting generalized eigenvalue problem.  While it may not be possible to represent general vectors in extremely large vector spaces, the inner products and matrix elements of eigenvectors can be computed using Monte Carlo simulations \cite{Frame:2017fah,Frame:2019jsw}, coupled cluster calculations \cite{Ekstrom:2019lss}, or some other many body method in order to solve the generalized eigenvalue problem.
EC has been used to deal with Monte Carlo sign oscillations \cite{Frame:2019jsw}, a resummation method for perturbation theory \cite{Demol:2019yjt,Demol:2020mzd}, and an accurate emulator for quantum systems \cite{Konig:2019adq}. More recently there have been a number of new developments, applications, and connections to other methods \cite{Ekstrom:2019lss,Furnstahl:2020abp,Bai:2021xok,Wesolowski:2021cni,Sarkar:2021sal,Yoshida:2021jbl,Melendez:2021lyq,Bonilla:2022rph,Melendez:2022kid}. The implementation of EC within an active learning framework was first discussed in Ref.~\cite{Eklind:2021car}.  However, one faces a computational bottleneck for large systems if the training process requires many repeated calculations of eigenvectors.  Here we instead use a fast estimate of the error function based upon the variance of the Hamiltonian. 

Let $H({\bf c})$ be a manifold of Hamiltonians where the dependence on the control parameters ${\bf c}$ is smooth.  Let $\ket{v({\bf c})}$ be the corresponding eigenvector of interest and $E({\bf c})$ be the corresponding energy eigenvalue.  The EC approximation consists of projecting $H({\bf c})$ onto the subspace spanned by the training eigenvectors $\{ \ket{v({\bf c}^{(i)})} \}$. By solving the generalized eigenvalue, we obtain the EC approximation to the eigenvector $\ket{\tilde v({\bf c})}$.  Throughout our discussion, we assume that all eigenvectors are unit normalized.  The corresponding approximate energy $\tilde E({\bf c})$ is equal to the expectation value $\braket{\tilde v({\bf c})| H({\bf c})|\tilde v({\bf c})}$.

The logarithm of the error is $\log \lVert \ket{\Delta v({\bf c})} \rVert$, where $\ket{\Delta v({\bf c})}=\ket{ v({\bf c})}-\ket{\tilde v({\bf c})}$.  Computing the error directly will be computationally too expensive for large systems, and so we will instead work with $\log F[\tilde v({\bf c}),H({\bf c})]$, where $F[\tilde v({\bf c}),H({\bf c})]$ is proportional to the square root of the variance of the Hamiltonian,
\begin{align}
    F[\tilde v({\bf c}),H({\bf c})] = \sqrt{ \frac{\braket{{\tilde v}({\bf c})| [ H({\bf c})-{\tilde E}({\bf c}) ]^2 | {\tilde v}({\bf c})}}{\braket{{\tilde v}({\bf c})| [ H({\bf c}) ]^2 | {\tilde v}({\bf c})}}}.
\end{align}
We note that $F[\tilde v({\bf c}),H({\bf c})]$ will be linearly proportional to $\lVert \ket{\Delta v({\bf c})} \rVert$ in the limit $\lVert \ket{\Delta v({\bf c})} \rVert \rightarrow 0$. Therefore $\log F[\tilde v({\bf c}),H({\bf c})]$ can be used as a surrogate for the logarithm of the error.

For Model 3 we consider the ground state of a system of four distinguishable particles with equal masses on a three-dimensional lattice with zero-range interactions.  We will work in lattice units where physical quantities are multiplied by the corresponding power of the lattice spacing to make dimensional combinations.  Furthermore, we set the particles masses to equal $1$ in lattice units.  We label the particles as $1,2,3,4$ and take the control parameters to be the six possible pairwise interactions, $c_{ij}$, with $i<j$. The lattice volume is a periodic cube of size $L^3=4^3$, and the corresponding Hamiltonian is a linear space with $262,144$ dimensions.  The details of the Hamiltonian can be found in the Supplemental Material. This model can be viewed as a generalization of the four two-component fermions with zero-range interactions considered in Ref.~\cite{Sarkar:2021sal,Bour:2011ad} or the Bose-Hubbard model considered in Ref.~\cite{Frame:2017fah}. 

We would like to study the appearance of interesting structures such as particle clustering \cite{Elhatisari:2017eno,Freer:2017gip} in the ground state wave function as a function of the six coupling parameters $c_{ij}$.  Some simple indicators of particle clustering are discussed in the Supplemental Material.  Such detailed multi-parameter studies are very difficult due to the number of repeated calculations necessary.  However, we now show that self-learning emulation with eigenvector continuation can make such studies fairly straightforward.

Since it is difficult to visualize data for all six parameters, we present results corresponding to one two-dimensional slice.  We set $c_{14}=c_{23}=c_{24}=c_{34}=-2.3475$ and use EC as an emulator for the ground state as a function of $c_{12}$ and $c_{13}$ over a square domain where each coefficient ranges from $-5$ to $5$.  We initialize the self-learning emulator with one random training point for $c_{12}$ and $c_{13}$.  When searching for new training points, we use the method of simulated annealing \cite{Pincus:1970} with an energy functional given by $-\log F[\tilde v({\bf c}),H({\bf c})]$.

\begin{figure*}[hbt]
     % \centering
\subfloat[\label{subfig:a}]{
\includegraphics[width=7.0cm]{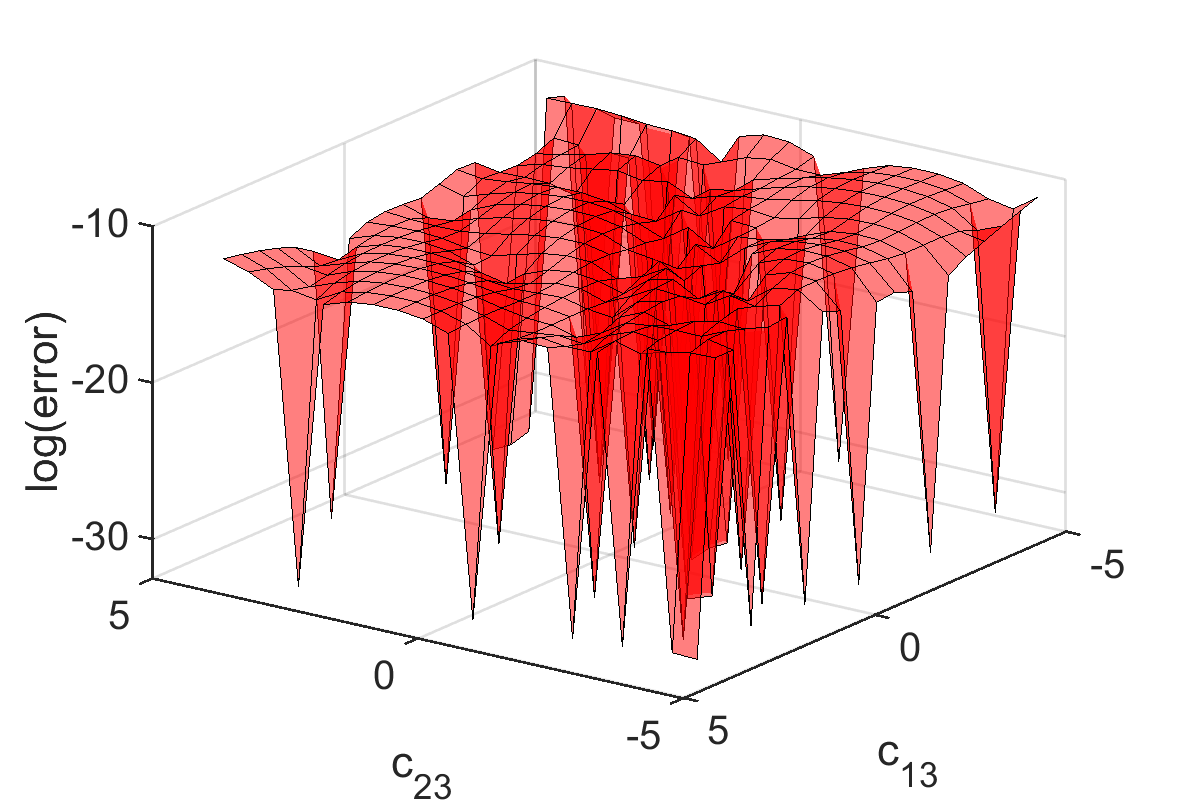}}
\subfloat[\label{subfig:b}]{
    \includegraphics[width=7.0cm]{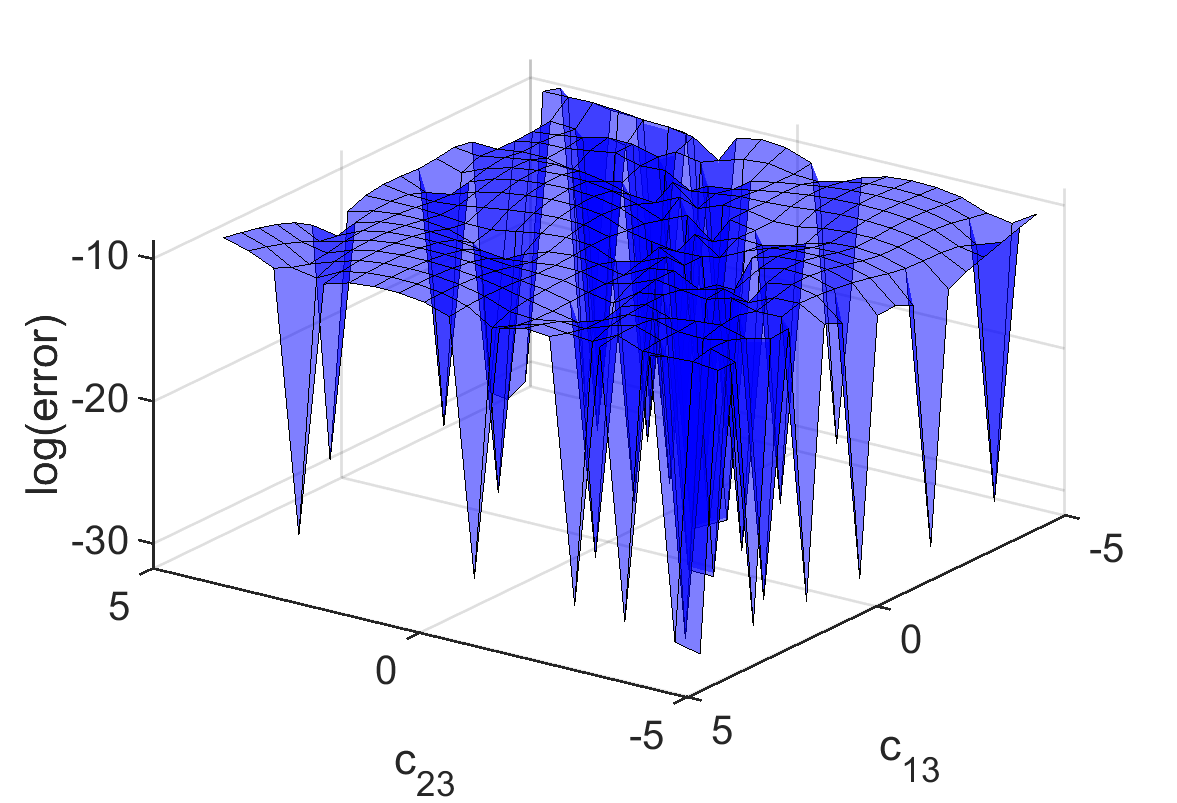}}
    \caption{Logarithm of the error in Model 3 after 40 iterations using self-learning EC. In panel (a) we show the logarithm of the actual error (red), and in panel (b) we show the logarithm of the estimated error (blue).}
     \label{Fig:error_plot_4particles}
\end{figure*}

\begin{figure}[hbt]
     % \centering
    \includegraphics[clip,width=\columnwidth]{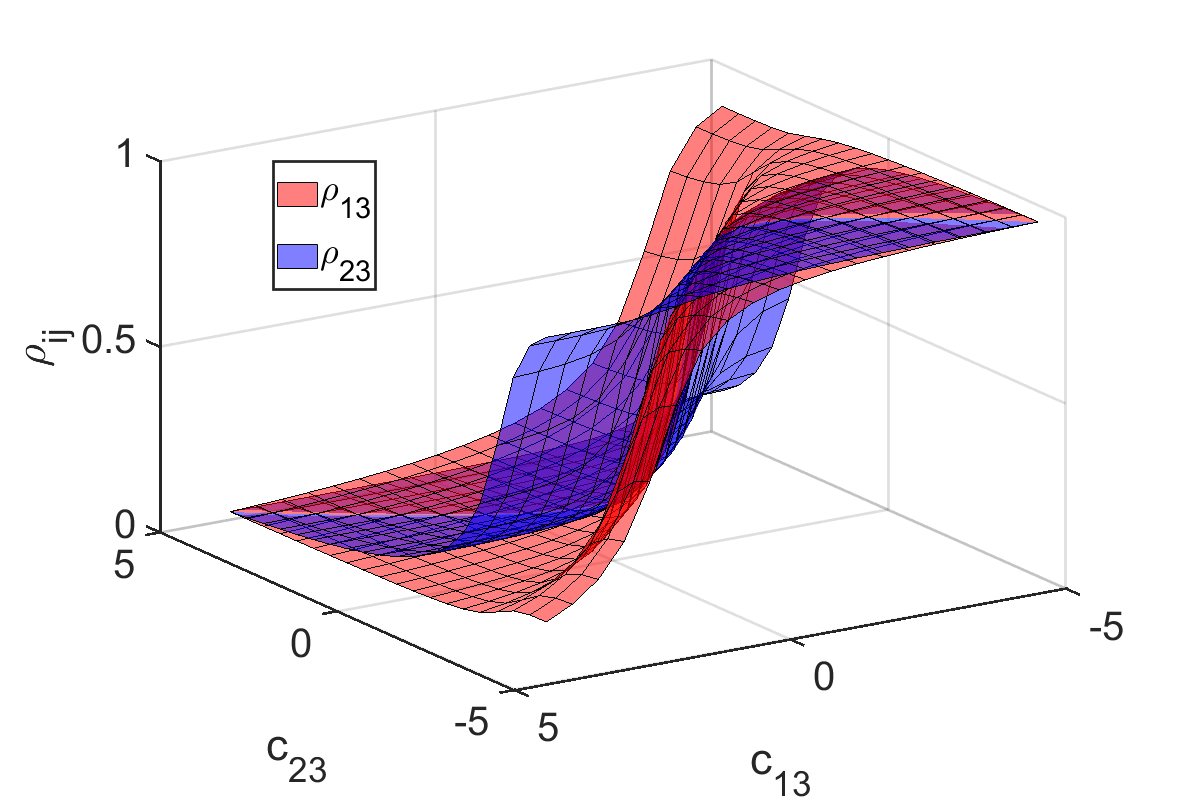}
    \caption{Plot of the two-particle clustering and short-range correlations in Model 3.  $\rho_{13}$ (red) measures the probability that particles $1$ and $3$ occupy the same lattice site, and the correlation function $\rho_{23}$ (blue) measures the probability that particles $2$ and $3$ occupy the same lattice site.}
     \label{Fig:2particle_cluster_prob}
\end{figure}

In Fig.~\ref{Fig:error_plot_4particles} we show the logarithm of the error obtained after 40 iterations. In panel (a) we show the logarithm of the actual error, and in panel (b) we show the logarithm of the estimated error.  As predicted in Eq.~(\ref{quick_error}), we see that the two plots are approximately the same up to a constant offset $A$, with $A \approx -2.3$.  The peak value of the actual error is $\lVert \ket{\Delta v({\bf c})} \rVert = 2\times 10^{-5}$. From the figure we see that the local maxima of the error reside along an approximately flat horizontal surface.  The flatness of this surface indicates that our self-learning emulator is performing as intended, with the training algorithm removing the peak error at each iteration.  We note that the distribution of training points is far from uniform. The region near the line $c_{12} + c_{13} = -1$ has a higher density of training points, indicating that the ground state wave function has a more complicated dependence on $c_{12}$ and $c_{13}$ in that location. 

The error scaling is exponential in the number of training points, $O(e^{-0.27N})$.  On a single Intel i7-9750H processor, direct calculation of the eigenvector and eigenvalue requires about $1.95$~s, whereas EC emulation with $41$ training points can be done in $0.013$~s. This corresponds to a raw speedup factor of $s_{\rm raw} \sim 150$.  
Using the self-learning emulator, we can now measure particle clustering and short-range correlations between pairs of particle in the ground state wave function for all values of $c_{12}$ and $c_{13}$.  In Fig.~\ref{Fig:error_plot_4particles} we show the short-range correlations for pairs of particles $1$ with $2$, and $1$ with $3$.  The correlation function $\rho_{12}$ measures the probability that particles $1$ and $2$ occupy the same lattice site, and the correlation function $\rho_{13}$ measures the probability that particles $1$ and $3$ occupy the same lattice site.  We see that $\rho_{12}$ is close to zero when $c_{12}$ is positive and rises to a peak of $1$ when $c_{12}$ is negative and increasing in magnitude.  Similarly, $\rho_{13}$ is close to zero when $c_{13}$ is positive and rises to a peak of $1$ when $c_{13}$ is negative and increasing in magnitude.  The change in structure is most prominent near the line $c_{12} + c_{13} = -1$, consistent with our emulator data on the selection of training points.  We have also studied the performance of the self-learning EC emulator when we vary all six control parameters $c_{ij}$ over the range from $-5$ to $0$. After $80$ iterations, the peak value of the error over the entire six-dimensional parameter space is $\lVert \ket{\Delta v({\bf c})} \rVert=4\times 10^{-3}$.

\paragraph*{Summary}   Self-learning emulation is a general approach that can be implemented with any emulator that faithfully reproduces the exact solution at selected training points.  They use a fast estimate for the error in the training process and perform full calculations only for the chosen new training points.  If needed, the difference between the estimated error and exact error can be corrected using machine learning.  If many evaluations are required, the computational advantage can grow as large as the raw speedup factor of the emulator, $s_{\rm raw}$, which can be several orders of magnitude or more.  Self-learning emulators are a highly efficient class of algorithms that offer both high speed and accuracy as well as a reliable estimate of the error.  

%\section*{Acknowledgement}
\paragraph*{Acknowledgement} We are grateful for discussions with E. Bonilla, J. Bonitati, R. Furnstahl, G. Given, P. Giuliani, K. Godbey, C. Hicks, M. Hjorth-Jensen, Da. Lee, J. Melendez, W. Nazarewicz, E. Ng, Z. Qian, J. Vary, J. Watkins, S. Wild, C. Yang, and X. Zhang.  We gratefully acknowledge funding by the U.S. Department of Energy (DE-SC0013365 and DE-SC0021152) and the Nuclear Computational Low-Energy
Initiative (NUCLEI) SciDAC-4 project (DE-SC0018083) as well as computational resources provided by the Oak Ridge Leadership Computing Facility through the INCITE award ``Ab-initio nuclear structure and nuclear reactions'', the Gauss Centre for Supercomputing e.V.
(www.gauss-centre.eu) for computing time on the GCS Supercomputer JUWELS at J{\"u}lich Supercomputing Centre (JSC), and Michigan State University. 

\bibliography{References}
\bibliographystyle{apsrev}
       
\beginsupplement
     
 \clearpage     

\onecolumngrid        

\section{Supplemental Material}

\subsection{Model 1 }

\begin{figure}[hbt]
     % \centering
    \includegraphics[width=8.4cm]{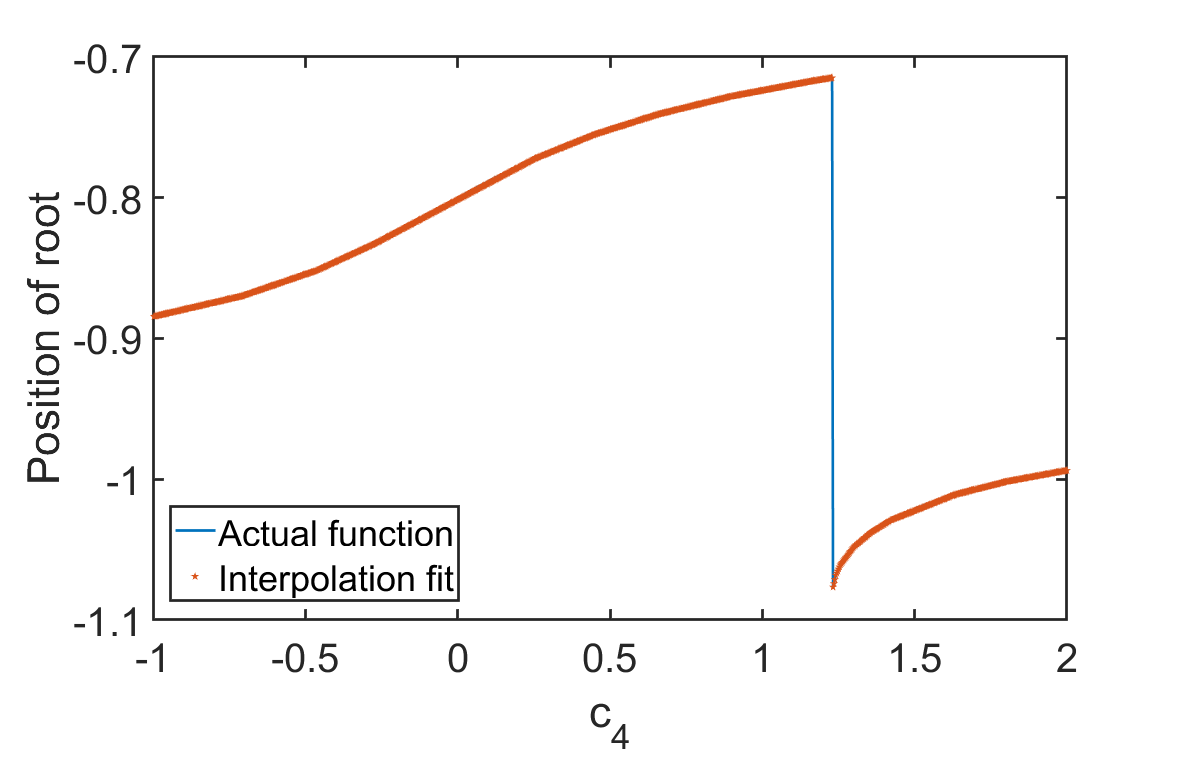}
    \caption{Plot of the lowest real solution to Eq.~(\ref{Eq:Sin_polynomial_supplem}) versus $c_4$. The self-learning emulator needs to take significantly more training points near the discontinuity at $c_4\approx1.232$.}
     \label{Fig:Sin_coefficient_polynomial_root_graph}
\end{figure}

In Model 1, we want the solution to the equation 
\begin{align}
    c_5 x^5 + c_4 x^4\sin (10x) + c_3 x^3 + c_2 x^2 + c_1 x + c_0 = 0 \label{Eq:Sin_polynomial_supplem}
\end{align}
where all the coefficients $c_i$ are real. We fix coefficients $c_5=c_3=c_2=c_1=c_0=1$, and we vary the coefficient $c_4$. With these choices, the lowest solution to the equation is shown in Fig.~\ref{Fig:Sin_coefficient_polynomial_root_graph}. We notice that the dependence of the solution on variable $c_4$ is non-trivial, and there is a discontinuity at $c_4\approx1.232$. As a result, the self-learning emulator needs to take significantly more training points near the discontinuity.

%\subsection{Speed-up factor of emulator}

\subsection{Dependence on initial training points}

\begin{figure}[hbt]
     % \centering
    \includegraphics[width=8.4cm]{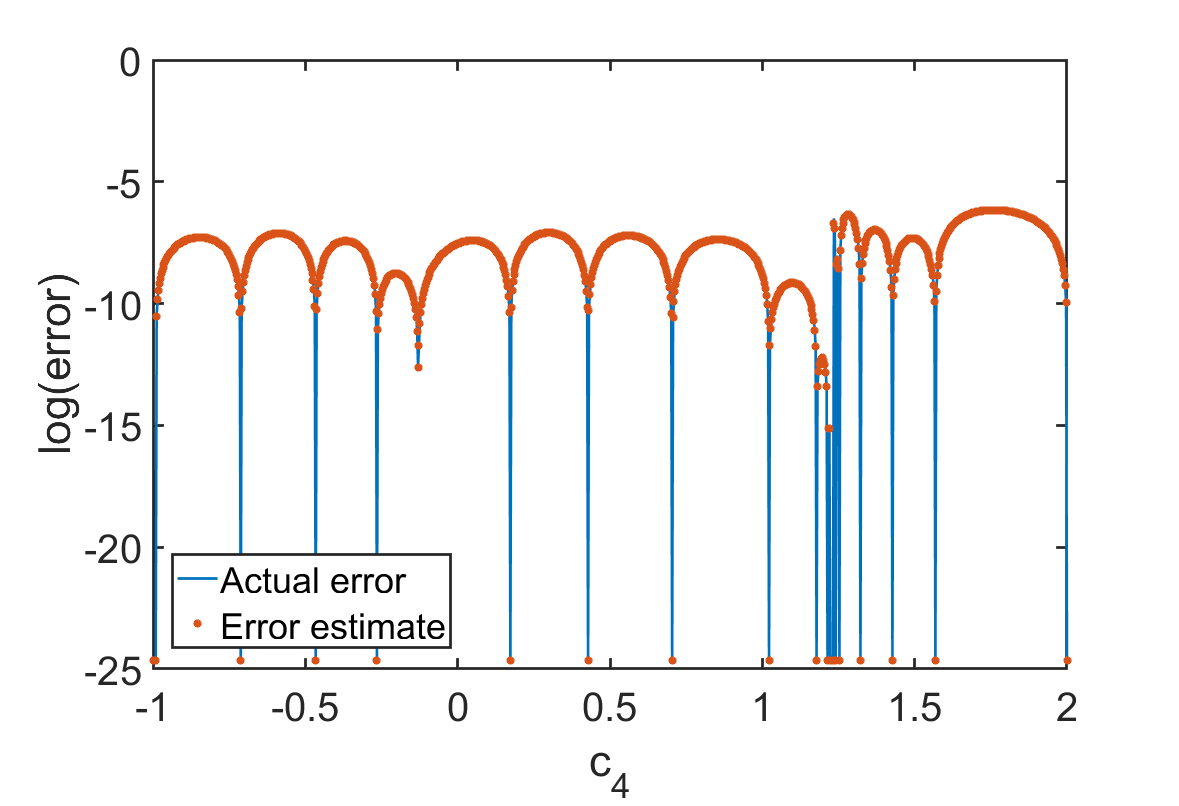}\\
    \caption{Logarithm of the actual error and error estimate for the cubic spline self-learning emulator in Model 1 after $20$ iterations when starting from training points $c_4 = -1.000, -0.997, -0.994$.}
    \label{Fig:spline_2_left}
\end{figure}

\begin{figure}[hbt]
     % \centering
    \includegraphics[width=8.4cm]{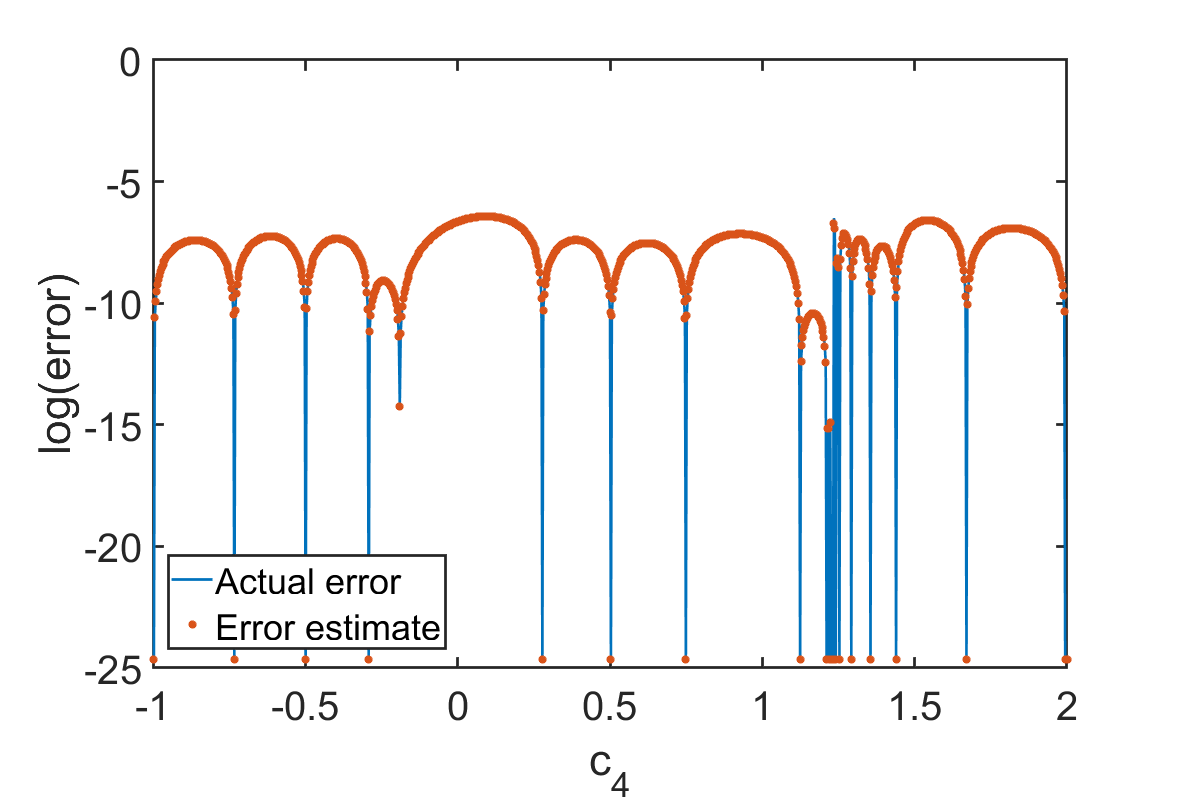}\\
    \caption{Logarithm of the actual error and error estimate for the cubic spline self-learning emulator in Model 1 after $20$ iterations when starting from training points $c_4 = 1.994, 1.997, 2.000$.}
    \label{Fig:spline_2_right}
\end{figure}

In this section we examine the performance of the self-learning emulator for Model 1 when starting from a poor choice of initial training points. In Fig.~\ref{Fig:spline_2_left}, we show the logarithm of the actual error and error estimate for the cubic spline self-learning emulator in Model 1 after $20$ iterations when starting from training points $c_4 = -1.000, -0.997, -0.994$.  In Fig.~\ref{Fig:spline_2_right}, we show the logarithm of the actual error and error estimate for the cubic spline self-learning emulator in Model 1 after $20$ iterations when starting from training points $c_4 = 1.994, 1.997, 2.000$.  We see that in both cases there is almost no loss of performance in comparison with Fig.~1 of the main text despite the poor choice of initial starting points.

\subsection{Error scaling}

\begin{figure}[hbt]
     % \centering
    \includegraphics[width=8.4cm]{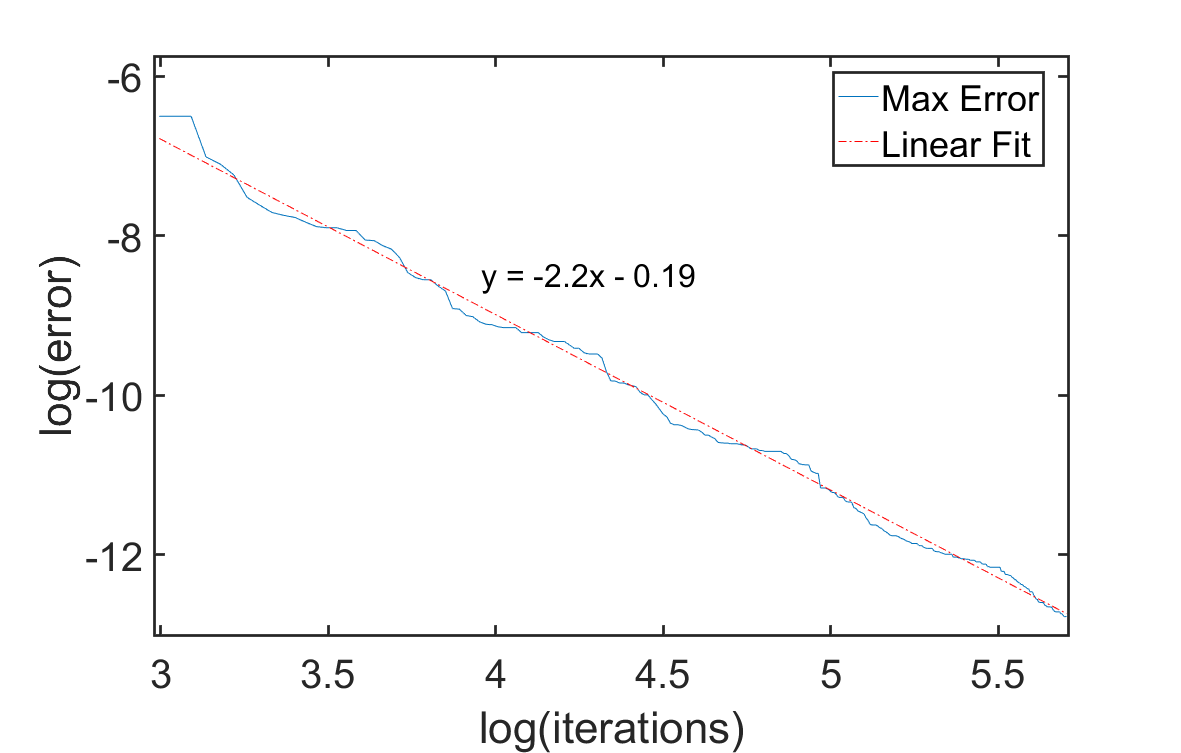}\\
    \caption{Natural spline emulator error scaling for Model 1.  We plot the logarithm of the error versus the logarithm of the number of iterations.}
    \label{Fig:Spline_error_convergence}
\end{figure}

If the solution is smoothly varying, we expect $O(N^{-4})$ error scaling for our self-learning natural spline emulator.  This is because the error of the cubic interpolation scales as the fourth power of the interval between training points.  However, this holds true only when the function is smooth and in the limit that $N$ is large.   For Model 1, however, the exact solution has a jump discontinuity, and so the power law scaling is slower.  Numerically, we find that the error is approximately $O(N^{-2.2})$.  We see this in Fig.~\ref{Fig:Spline_error_convergence}, where the slope of the graph is $-2.2$.

\begin{figure}[hbt]
     % \centering
    \includegraphics[width=8.4cm]{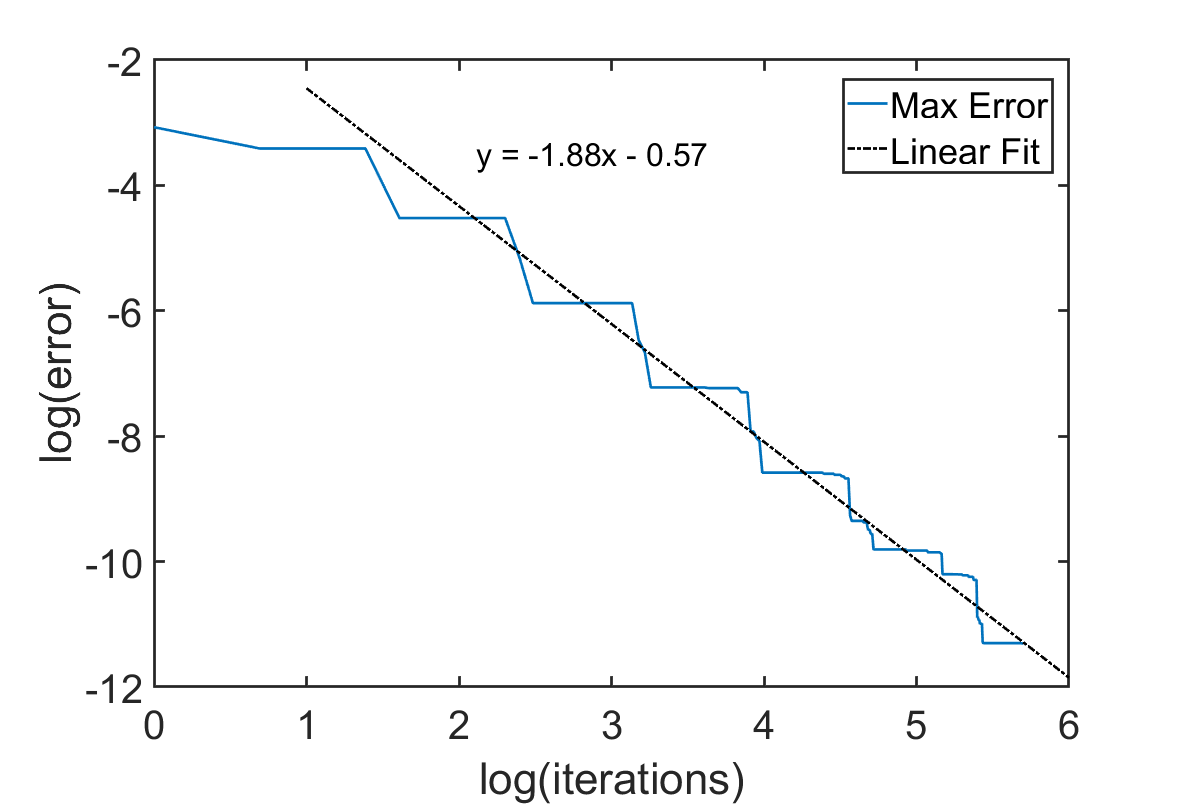}\\
    \caption{Natural spline emulator error scaling for Model 2.  We plot the logarithm of the error versus the logarithm of the number of iterations.}
    \label{Fig:Spline_ODE_error_convergence}
\end{figure}

\begin{figure}[hbt]
     % \centering
    \includegraphics[width=8.4cm]{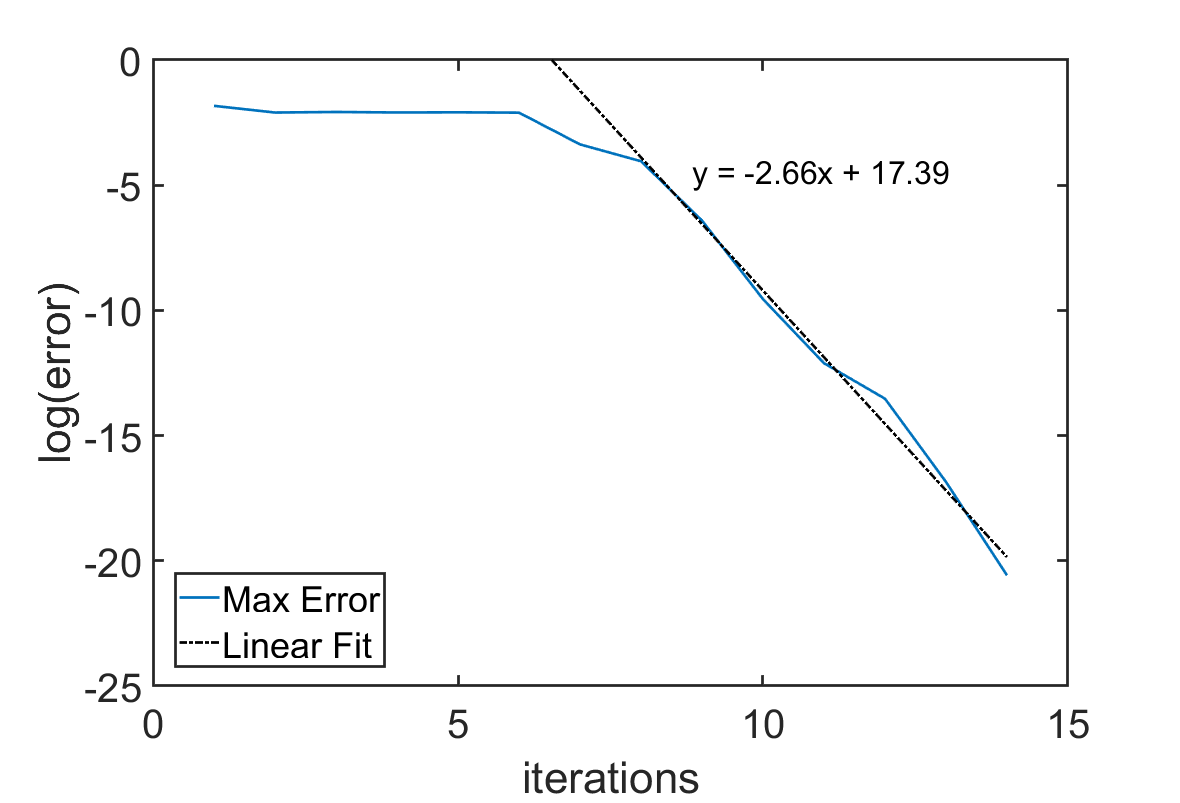}\\
    \caption{Reduced basis method error scaling for Model 2.  We plot the logarithm of the error versus the number of iterations.}
    \label{Fig:Reduced_basis_error_convergence}
\end{figure}

In Model 2, the solution is smoothly varying function.  However it seems that have not yet reached the asymptotic scaling large $N$ limit, and the error scaling is approximately $O(N^{-1.88})$.  This can be seen from the $-1.88$ slope in Fig.~\ref{Fig:Spline_ODE_error_convergence}. In contrast, the reduced basis emulator has exponentially fast error scaling.  This is because the reduced basis emulator is itself a smooth function.  We can view the addition of training points as matching more derivatives of the smooth emulator to derivatives of the smooth exact solution.  The error scaling is therefore similar to the error scaling of a convergent power series.
In Fig.~\ref{Fig:Reduced_basis_error_convergence} we show the error scaling for the reduced basis emulator for Model 2.  We see that the error scaling is $O(e^{-2.66N})$.

\begin{figure}[hbt]
     % \centering
    \includegraphics[width=8.4cm]{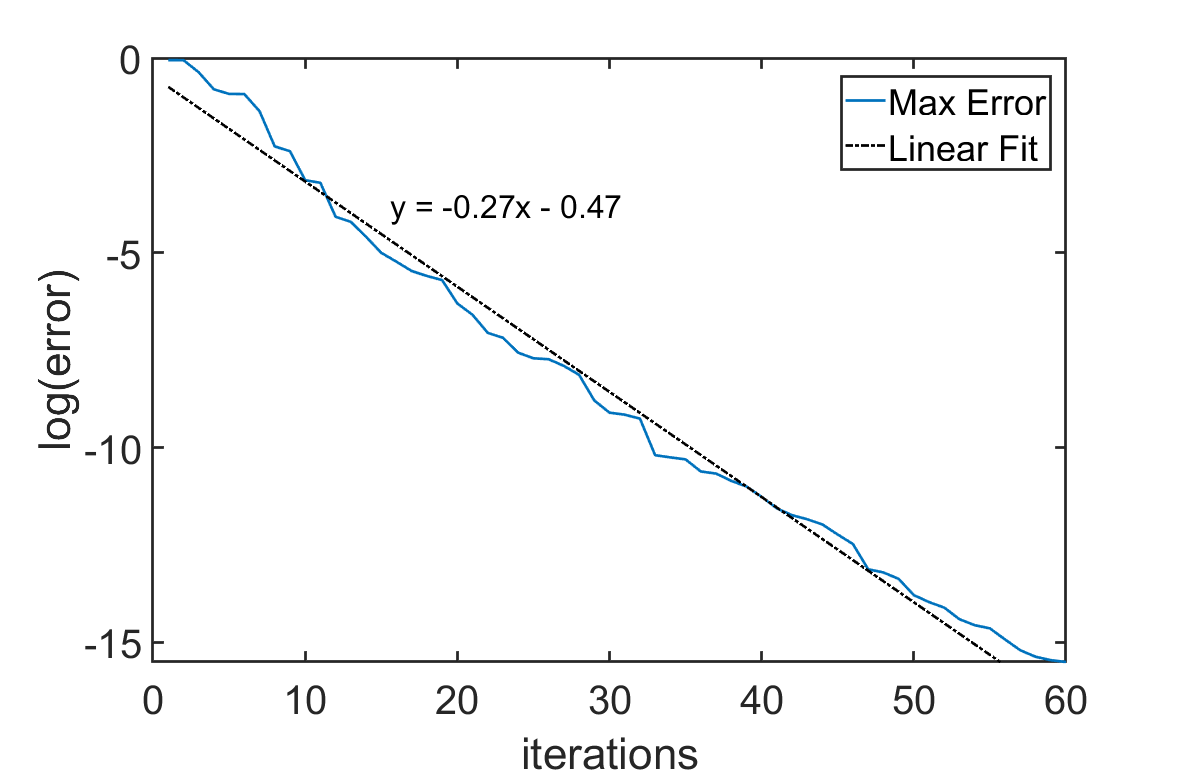}\\
    \caption{Eigenvector continuation emulator error scaling for Model 3.  We plot the logarithm of the error versus the number of iterations.}
    \label{Fig:EC_error_convergence}
\end{figure}

For the eigenvector continuation emulator in Model 3, we again expect exponential error scaling because both the emulator and exact solution are both smoothly varying functions.  In Fig.~\ref{Fig:EC_error_convergence} we show the error scaling for the eigenvector continuation emulator in Model 3.   We see that the error scaling is $O(e^{-0.27N})$.

\subsection{Geometrical picture of eigenvector continuation error}

We will present a geometrical picture of eigenvector continuation (EC) error as well as some additional insight into the error estimate that appears in Eq.~(3) of the main text.  % This equation number is written explicitly and needs to be updated if the equation number changes in the main text. 
We consider a Hamiltonian manifold $H({\bf c})$ that depends on the control parameters ${\bf c}$.  We write $\ket{v({\bf c})}$ for the eigenvector of interest and $E(\bf c)$ for the corresponding energy eigenvalue.  Suppose we know the eigenvectors at $M$ different training points, $\{{\bf c}^{(1)},\cdots,{\bf c}^{(M)}\}$.  We label the set of $M$ training eigenvectors as $S_M = \{\ket{v({\bf c}^{(1)})},\cdots,\ket{v({\bf c}^{(M)})}\}$.  Let us define the norm matrix ${\cal N}(S_M)$ as
\begin{align}
	\begin{bmatrix}
	\braket{v({\bf c}^{(1)})|v({\bf c}^{(1)})} &\cdots & \braket{v({\bf c}^{(1)})|v({\bf c}^{(M)})}\\
	\vdots &\ddots &\vdots\\
    \braket{v({\bf c}^{(M)})|v({\bf c}^{(1)})} &\cdots & \braket{v({\bf c}^{(M)})|v({\bf c}^{(M)})}\\
	\end{bmatrix},
\end{align}
and let $\Omega^2(S_M)$ be the determinant of ${\cal N}(S_M)$.  Then $\Omega^2(S_M)$ corresponds to the square of the volume of the $M$-dimensional parallelopiped defined by the vectors in the set $S_M$. If all the eigenvectors are normalized, then the maximum possible volume is 1, which is attained when all the eigenvectors are orthogonal.

Let us now consider selecting the next training point, ${\bf c}_{M+1}$. Let $P$ be the projection operator onto the linear span of $S_{M}$, and let $Q$ be the orthogonal complement so that $Q=1-P$. Suppose we now expand our training set $S_{M}$ by adding another training vector $\ket{v({\bf c})}$ to form $S_{M+1}$. Let us define the perpendicular projection vector $\ket{v_{\perp}({\bf c})}$ as
\begin{align}
    \ket{v_{\perp}({\bf c})} = Q\ket{v({\bf c})}.
\end{align}
Since $\Omega^2(S_M)$ is the squared volume of the parallelopiped defined by the vectors in $S_M$ and $\Omega^2(S_{M+1})$ is the squared volume of the parallelopiped defined by the vectors in $S_{M+1}$, it follows that the ratio $\Omega^2(S_{M+1})$ to $\Omega^2(S_{M})$ is given by the squared norm of $\ket{v_{\perp}({\bf c})}$,
\begin{align}
    \frac{\Omega^2(S_{M+1})}{\Omega^2(S_{M})} = \braket{v_{\perp}({\bf c})|v_{\perp}({\bf c})}.
    \label{Eq:determinant_ratio2}
\end{align}

Let us define the projections of $H$ onto $P$ and $Q$ subspaces as
\begin{align}
    H^{P}({\bf c}) = PH({\bf c})P,\qquad H^{Q}({\bf c}) = QH({\bf c})Q.
\end{align}
The EC approximation is nothing more than the approximation of $\ket{v({\bf c})}$ by some eigenvector of $H^P({\bf c})$, which we denote as $\ket{v^P({\bf c})}$.  Let the corresponding energy be labelled $E^P({\bf c})$ so that
\begin{align}
    H^{P}({\bf c}) \ket{v^{P}({\bf c})} = E^{P}({\bf c})\ket{v^{P}({\bf c})}.
\end{align}
We also label the eigenvectors of $H^{Q}({\bf c})$ contained in the orthogonal complement $Q$ as,
\begin{align}
    H^{Q}({\bf c}) \ket{v^{Q}_j({\bf c})} = E^{Q}({\bf c})\ket{v^{Q}_j({\bf c})}.
\end{align}

When the difference between the exact eigenvector and the eigenvector continuation approximation of the eigenvector is small, we can use first order perturbation theory to write
\begin{align}
    \ket{v({\bf c})} \approx \ket{v^{P}({\bf c})} + \sum_j \frac{\braket{v^{Q}_j({\bf c})|H({\bf c})|v^{P}({\bf c})}}{E^{P}({\bf c})-E^{Q}_j({\bf c})} \ket{v^{Q}_j({\bf c})}.
\end{align}
To first order in perturbation theory, the residual vector is just $\ket{v_\perp({\bf c})} \approx \ket{v({\bf c})} - \ket{v^{P}({\bf c})}$.  We therefore have
\begin{align}
    \ket{v_\perp({\bf c})} \approx \sum_j \frac{\braket{v^{Q}_j({\bf c})|H({\bf c})|v^{P}({\bf c})}}{E^{P}({\bf c})-E^{Q}_j({\bf c})} \ket{v^{Q}_j({\bf c})}
\end{align}
If we now combine with Eq.~(\ref{Eq:determinant_ratio2}), we get
\begin{align}
      \frac{\Omega^2(S_{M+1})}{\Omega^2(S_{M})} = \lVert \ket{v_{\perp}({\bf c})} \rVert^2 
      = \sum_j \frac{\braket{v^{P}({\bf c})|H({\bf c})|v^{Q}_j({\bf c})}\braket{v^{Q}_j({\bf c})|H({\bf c})|v^{P}({\bf c})}}{[E^{P}({\bf c})-E^{Q}_j({\bf c})]^2}.\label{Eq:determinant_ratio3}
\end{align}
We can now connect this result with the error or loss function in the main text.  The second part of the equation gives an expression for the error term $\lVert \ket{v_{\perp}({\bf c})} \rVert$ using first-order perturbation theory, and the first part of the equation is a geometrical interpretation of the error term as the ratio of the squared volumes, $\Omega^2(S_{M+1})$ to $\Omega^2(S_{M})$. Taking the logarithm of the square root, we get
\begin{align}
     \log \lVert \ket{v_{\perp}({\bf c})} \rVert = 
    \frac{1}{2}\log\sum_j \frac{\braket{v^{P}({\bf c})|H({\bf c})|v^{Q}_j({\bf c})}\braket{v^{Q}_j({\bf c})|H({\bf c})|v^{P}({\bf c})}}{[E^{P}({\bf c})-E^{Q}_j({\bf c})]^2}.\label{Eq:Log_error}
\end{align}
The term in the numerator, 
\begin{align}
    \braket{v^{P}({\bf c})|H({\bf c})|v^{Q}_j({\bf c})}\braket{v^{Q}_j({\bf c})|H({\bf c})|v^{P}({\bf c})},
\end{align} 
will go to zero at each of the training points, causing large variations in the logarithm of the error as we add more and more training points.  In contrast, the term in the denominator, $[E^{P}({\bf c})-E^{Q}_j({\bf c})]^2$, will be smooth as a function of $c$. Similarly, $\braket{v^{P}({\bf c})|[H({\bf c})]^2|v^{P}({\bf c})}$ will also be a smooth function of ${\bf c}$.  We can write
\begin{align}
    \frac{1}{2}\log\sum_j \frac{\braket{v^{P}({\bf c})|H({\bf c})|v^{Q}_j({\bf c})}\braket{v^{Q}_j({\bf c})|H({\bf c})|v^{P}({\bf c})}}{[E^{P}({\bf c})-E^{Q}_j({\bf c})]^2} =
    \frac{1}{2}\log\sum_j \frac{\braket{v^{P}({\bf c})|H({\bf c})|v^{Q}_j({\bf c})}\braket{v^{Q}_j({\bf c})|H({\bf c})|v^{P}({\bf c})}}{\braket{v^{P}({\bf c})|[H({\bf c})]^2|v^{P}({\bf c})}} + A + B({\bf c}),
\end{align}
where $A$ is a constant and $B({\bf c})$ averages to zero over the entire domain of ${\bf c}$.  While the function $B({\bf c})$ is unknown, it will be dominated by the large variations in the logarithm of the error as more and more training points are added.  We note that
\begin{align}
    \sum_j & \frac{\braket{v^{P}({\bf c})|H({\bf c})|v^{Q}_j({\bf c})}\braket{v^{Q}_j({\bf c})|H({\bf c})|v^{P}({\bf c})}}{\braket{v^{P}({\bf c})|[H({\bf c})]^2|v^{P}({\bf c})}} =\frac{\braket{v^{P}({\bf c})|H({\bf c})(1-P)(1-P)H({\bf c})|v^{P}({\bf c})}}{\braket{v^{P}({\bf c})|[H({\bf c})]^2|v^{P}({\bf c})}} \nonumber \\
    & =\frac{\braket{v^{P}({\bf c})|[H({\bf c}) - H^{P}({\bf c})]^2|v^{P}({\bf c})}}{\braket{v^{P}({\bf c})|[H({\bf c})]^2|v^{P}({\bf c})}} 
    =\frac{\braket{v^{P}({\bf c})|[H({\bf c}) - E^{P}({\bf c})]^2|v^{P}({\bf c})}}{\braket{v^{P}({\bf c})|[H({\bf c})]^2|v^{P}({\bf c})}}. 
\end{align}
We therefore arrive at the variance error estimate used in the main text,
\begin{align}
    \log \lVert \ket{v_{\perp}({\bf c})} \rVert = \frac{1}{2}\log \frac{\braket{v^{P}({\bf c})|[H({\bf c}) - E^{P}({\bf c})]^2|v^{P}({\bf c})}}{\braket{v^{P}({\bf c})|[H({\bf c})]^2|v^{P}({\bf c})}} + A + B({\bf c}).
\end{align}

\subsection{Model 3 Hamiltonian}
Model 3 describes four-distinguishable particles with equal masses $m$ on a three-dimensional lattice with pairwise point interactions with coefficients $c_{ij}$ for each pair $i<j$.  We use lattice units where physical quantities are multiplied by powers of the spatial lattice spacing to make the combinations dimensionless.  We take the common mass $m$ to equal $1$ in lattice units. We let ${\bf n}$ denote the spatial lattice points on our three dimensional $L^3$ periodic lattice. Let the lattice annihilation and creation operators for particle $i$ be written as $a_{i}({\bf n})$ and $a^\dagger_{i}({\bf n})$ respectively.  The free non-relativistic lattice Hamiltonian has the form
\begin{align}
    H_{\text{free}} = \frac{3}{m}\sum_{i=1,2,3,4}\sum_{{\bf n}}
    a_{i}^\dagger({\bf n})a_{i}({\bf n})
    -&\frac{1}{2m}\sum_{i=1,2,3,4}\sum_{{\bf \hat{l}}={\bf \hat{1}},{\bf \hat{2}},{\bf \hat{3}}}
    \sum_{{\bf n}}
    a_{i}^\dagger({\bf n})\Big[a_{i}({\bf n}+{\bf \hat{l}})+a_{i}({\bf n}-{\bf \hat{l}})\Big].
\end{align}
We add to the free Hamiltonian single-site contact interactions, and the resulting Hamiltonian then has the form
\begin{equation}
    H = H_{\text{free}} + \sum_{i<j}\sum_{{\bf n}}c_{ij}\rho_{i}({\bf n})\rho_{j}({\bf n}),
\end{equation}
where $\rho_{i}({\bf n})$ is the density operator for particle $i$,
\begin{align}
    \rho_{i}({\bf n}) &= a_{i}^\dagger({\bf n})a_{i}({\bf n}).
\end{align}

For calculations discussed in this work, we use a basis of position eigenstates on the lattice.  As noted in Ref.~\cite{Elhatisari:2017eno}, we can determine the formation of particle clusters by measuring the expectation values of products of local density operators.  For example, $\rho_{ij}({\bf n})=\rho_{i}({\bf n})\rho_{j}({\bf n})$ can serve as an indicator of two-particle clusters, $\rho_{ijk}({\bf n})=\rho_{i}({\bf n})\rho_{j}({\bf n})\rho_{k}({\bf n})$ for three-particle clusters, and $\rho_{ijkl}({\bf n})=\rho_{i}({\bf n})\rho_{j}({\bf n})\rho_{k}({\bf n})\rho_{l}({\bf n})$ for a four-particle cluster.
\end{document}